\documentstyle[seceq,supplement]{ptptex}

\def\k{\mbox{\boldmath$k$}}
\def\x{\mbox{\boldmath$x$}}

\def\k{\mbox{\boldmath$k$}}
\def\x{\mbox{\boldmath$x$}}

\title{
 Cosmological Perturbations in Brane World
}
\subtitle{Brane view v.s. Bulk view}

\author{
Jiro {\sc Soda}\footnote{E-mail: jiro@phys.h.kyoto-u.ac.jp} and 
Kazuya {\sc Koyama}$^{}$\footnote{E-mail: kazuya@utap.phys.s.u-tokyo.ac.jp}
}

\inst{
 $^*$Department of Fundamental Sciences, FIHS, Kyoto University, Kyoto 606-8501
\\
$^{**}$Department of Physics, The University of Tokyo, Bunkyo-ku, 113-0033}

\recdate{
}

\abst{%
 First, we will study the cosmological perturbations from the brane point of 
 view. It turns out that two types of the extra data are necessary to know
  the evolution of the system. To fix these data, the analysis of the bulk 
  is needed. So, we have solved  equations of motion for the bulk gravity
  and determined the extra data. We would like to stress that,
  both analysis take complementary roles to achieve this goal. 
}

\begin{document}

\maketitle

\section{Introduction }
Recently, there has been much interest in the brane world scenario.~\cite{RS}
The action for the ($Z_2$ symmetric) brane world is given by
\begin{equation}
S= \frac{1}{2 \kappa^2}\int d^5 x \sqrt{-g}
\left(
{\cal R}^5 +  \frac{12}{l^2} \right)
- \sigma \int d^4 x \sqrt{-g_{brane}}
+ \int d^4 x \sqrt{-g_{brane}} {\cal L}_{matter}
\end{equation}
where  $l, \kappa^2, \sigma $ are the curvature radius of the AdS spacetime, 
  the gravitation constant in the 5D spacetime, 
  and  the brane tension, respectively. Here we assume the relation
   $\kappa^2 \sigma=6/l$.
  
 The importance of understanding the cosmological perturbations in the brane 
 world is widely recognized.~\cite{KJ,Muko,kodama,Bruck,Lan,Mar,Der,Lan2} 
 Needless to say, the data on the brane are not sufficient to determine the
 evolution of the system. The analysis of the equations for the bulk fields
 is crucial.
 In sec.2, we will demonstrate this fact using the 
 example of the homogeneous cosmology. In sec.3, the junction condition for 
 the perturbed spacetime is presented. In sec.4, we will identify two types of
 the extra data from the analysis on the brane. In sec.5, we will solve the
 5-dimensional problem using the transformation method. In the final section,
 the extra data are fixed by making use of the both pictures complementarily.

\section{Brane World Cosmology}

The background metric for the brane world cosmology is
\begin{equation}
ds^2=e^{2 \beta(y,t)}(dy^2-dt^2)+ e^{2 \alpha(y,t)}
\delta_{ij} dx^i dx^j
\end{equation}
and then the induced metric on the brane becomes 
$
ds^2 =-dt^2+e^{2 \alpha_0(t)} \delta_{ij} dx^i dx^j
$. 
The energy momentum tensor of the matter is taken as 
$
 T^M_N= diag(0,-\rho,p,p,p) \delta(y)
$. 
We will denote the power series expansion near the brane 
as 
$
\alpha(y,t)=\alpha_0(t)+ \alpha_1(t) \vert y \vert  + 
\frac{\alpha_2(t)}{2} y^2+ \cdots $. 
Then the  junction conditions are obtained as
\begin{equation}
\alpha_1(t) = - \frac{1}{l}   
- \frac{\kappa^2 \rho(t)}{6} \ ,  \quad
\beta_1(t) =  - \frac{1}{l}+ 
\frac{\kappa^2 \rho(t)}{3} + \frac{ \kappa^2 p(t)}{2} \ ,
\end{equation}
where we set $e^{\beta_0(t)}=1$.

\subsection{ the view from the brane}
  The effective Friedmann equation holds on the brane;~\cite{B,F}
\begin{equation}
\dot{\alpha}_0^2 = \frac{8\pi G_4}{3} \rho + 
\frac{\kappa^4 \rho^2}{36} + e^{-4 \alpha_0} C_0  \ ,
\end{equation}
where $  8 \pi G_4=\kappa^2/l $ and $C_0$ is the constant of the integration. 
We also have the conservation law 
$ \dot{\rho}+ 3 \dot{\alpha}_0 (\rho+p)=0 $. Hence,
at low energies $\rho/\sigma \sim \kappa^{-2} l \rho \ll 1$, we obtain 
 the conventional equations when $C_0 =0$. 

 On the brane, we do not have any criterion to determine the
  extra data $C_0$. To determine it, 
 we have to resort to the analysis of the bulk. 

\subsection{ the view from the bulk}

A natural boundary condition to specify the bulk geometry is to impose the 
 zero Weyl curvature, i.e.  Anti-deSitter bulk.
Starting with the AdS spacetime
\begin{equation}
ds^2= \left( \frac{l}{z} \right)^2
\left( dz^2 -d \tau^2 + \delta_{ij} dx^i dx^j \right)
\end{equation}
and making the coordinate transformation: 
$
 z=l \left(f(u)-g(v)\right), \quad \tau=l \left(f(u)+g(v) \right)
$  with  $u=(t-y)/l,v=(t+y)/l$, 
we have
\begin{equation}
e^{2 \beta(y,t)} =  4 \frac{f'(u)g'(v)}{(f(u)-g(v))^2}, \qquad
e^{2 \alpha(y,t)} = \frac{1}{(f(u)-g(v))^2}  \ .
\end{equation}
Thus we have the scale factor of the brane as 
 $e^{\alpha_0(t)} =(f(t/l)-g(t/l))^{-1}$.
Using the junction condition 
\begin{equation}
\alpha_1 = \frac{1}{l} \left( \frac{f(t/l)'+g(t/l)'}{f(t/l)-g(t/l)} 
\right)
 =-\frac{1}{l}-
\frac{\kappa^2 \rho}{6}       
\end{equation}
and the gauge condition
\begin{equation}
e^{2 \beta_0} = 4 \frac{f(t/l)' g(t/l)'} {(f(t/l)-g(t/l))^2}=1 \ ,
\end{equation}
we obtain
\begin{equation}
\dot{\alpha}_0^2 = \frac{1}{l^2} 
\left( \frac{f(t/l)'-g(t/l)'}{f(t/l)-g(t/l)} \right)^2
 = \alpha_1^2-1/l^2  
 =\frac{\kappa^2}{3 l}\rho+\frac{\kappa^4 \rho^2}{36} 
                            \ .
\end{equation}
Thus, by specifying the bulk as the Anti-de Sitter spacetime,
 the extra data $C_0$ is determined as zero.

\section{Cosmological Perturbations: Junction conditions}

Let us consider the perturbed 5D AdS spacetime in the 
Gaussian Normal coordinate:
\begin{eqnarray}
ds^2 &=& e^{2 \beta(y,t)} \left((1+2 N) dy^2 -(1+2 \Phi)
d t^2  \right)   \nonumber\\
&+&
e^{2 \alpha(y,t)}
\left(\left((1 - 2 \Psi) \delta_{ij}+ 2 E_{,ij}\right) 
dx^i dx^j+ 2 B_{,i} dx^i d t  \right) \ .
\end{eqnarray}
The Einstein equation 
$
\delta G^M_N = 
\kappa^2 e^{- \beta} (\delta T^M_N - N T^{M}_{N})
$
and the 5D energy-momentum tensor:
\begin{equation}
\delta T^M_N = 
\left(
\begin{array}{ccc}
0 & 0 & 0 \\
0 & -\delta \rho & -(\rho+p)e^{\alpha_0} v_{,i} \\
0 & (\rho+p)e^{-\alpha_0} v_{,i}  & \delta p \: \delta_{ij}\\
\end{array}
\right) \: \delta (y)
\end{equation}
gives the junction conditions (notice that the brane is located at $y=0$)
\begin{eqnarray}
\Psi_1 &=& -\alpha_1 N_0 +\frac{1}{6} \kappa^2 
\delta \rho   \ , \nonumber\\
\Phi_1 &=&  \beta_1 N_0 + \kappa^2 
\left(\frac{\delta \rho}{3}+
\frac{\delta p}{2} \right) \ ,\nonumber\\
B_1 &=&  -2 (\beta_1-\alpha_1) e^{-\alpha_0} v \ , \nonumber\\
E_1 &=&  0    \ .
\end{eqnarray}
Here we have also taken $B_0(y=0,t,x^i)=E_0(y=0,t,x^i)=0$ gauge.


\section{ Cosmological Perturbations: the view from the brane}

With the perturbed metric on the brane
\begin{equation}
ds_{brane}=-(1+2 \Phi_0) dt^2+e^{2\alpha_0(t)}
(1-2 \Psi_0) \delta_{ij} dx^i dx^j   \ ,
\end{equation}
the equations on the brane become
\begin{eqnarray}
\ddot{\Psi}_0+4 \dot{\alpha}_0 \dot{\Psi}_0 +\dot{\alpha}_0 \dot{\Phi}_0
+2(\ddot{\alpha}_0+2 \dot{\alpha}_0^2) \Phi_0 
- \frac{1}{3} e^{-2 \alpha_0} (2 \nabla^2 \Psi_0 -\nabla^2 \Phi_0)
\nonumber\\
=
\frac{\kappa^2}{3} \left(\frac{\beta_1}{2} 
\delta \rho -\frac{3 \alpha_1}{2} \delta p \right)   \ ,
\end{eqnarray}
\begin{equation}
\dot{\delta \rho} =(\rho+p)(3 \dot{\Psi}_0+ e^{-\alpha_0} 
\nabla^2 v)
-3 \dot{\alpha}_0 
\left( \delta \rho+ \delta p \right)  \ , 
\end{equation}
\begin{equation}
\left( (\rho+p) e^{\alpha_0} v\right)^{\cdot}
= -3 \dot{\alpha}_0 (\rho+p) e^{\alpha_0} v + \delta p 
+(\rho+p) \Phi_0   \ .
\end{equation}
In the conventional 4D cosmology, in addition to these equations, 
we have the equation $ \Phi_0-\Psi_0 = 0 $ for matter with 
no anisotropic stress and the Hamiltonian constraint equation. 
The interesting point is that, even in the case of $\Psi_0=\Phi_0$, 
the corrections to the matter perturbations can exist. 
In order to see this, we regard them as equations
 for $\delta \rho, \delta p $ and $v$. 
The solutions can be obtained by gradient expansion.
 They are divided into two parts;
 the homogeneous solution written by $\chi$ and 
 the particular solution.

The solutions are 
\begin{equation}
-\frac{\kappa^2 \alpha_1}{2} \delta \rho
     = -3 (\dot{\alpha}_0 \dot{\Psi}_0 + \dot{\alpha}_0^2 \Phi_0)
       +e^{-2 \alpha_0} \nabla^2 \Psi_0  
       + \frac{\kappa^2}{2 l}\delta \rho_{\chi} 
       + {\cal O}(\nabla^4 \Psi_0 )    
\end{equation}
\begin{eqnarray}
  && -\frac{\kappa^2 \alpha_1}{2} \delta p
= \ddot{\Psi}_0+\left(3 \dot{\alpha}_0-\frac{\dot{\alpha_0} 
\ddot{\alpha}_0}{\alpha_1^2}\right) \dot{\Psi}_0+
\dot{\alpha}_0 \dot{\Phi}_0 
  +\left( 2 \ddot{\alpha}_0
-\frac{\dot{\alpha}_0^2 \ddot{\alpha}_0}{\alpha_1^2} +3 \dot{\alpha}_0^2
\right) \Phi_0    \nonumber\\ 
&&\qquad + \frac{1}{3}e^{-2 \alpha_0}\nabla^2 \Phi_0-\frac{1}{3}
\left(1-\frac{\ddot{\alpha}_0}{\alpha_1^2} \right)
e^{-2 \alpha_0}\nabla^2 \Psi_0 
   + \frac{\kappa^2}{2 l}
\delta p_{\chi}
   +  {\cal O}(\nabla^4 \Psi_0 )   
\end{eqnarray} 
\begin{eqnarray}  
-\frac{\kappa^2 \alpha_1}{2} (\rho+p) e^{\alpha_0}v &=&
\dot{\Psi}_0 +\dot{\alpha}_0 \Phi_0   \nonumber \\
&& +\frac{1}{3} \alpha_1 e^{-3 \alpha_0} \int dt'
 e^{\alpha_0} \alpha_1^{-1} \left( \nabla^2 \Phi_0-\left(1-
\frac{\ddot{\alpha}_0}{\alpha_1^2} \right) \nabla^2 \Psi_0
\right)     \nonumber \\
&&  + \frac{\kappa^2}{2 l} (\rho+p)e^{\alpha_0} v_{\chi}
     + {\cal O}(\nabla^4 \Psi_0 )  \ .
\end{eqnarray}
Homogeneous solutions $\delta \rho_{\chi}$, $\delta p_{\chi}$ and
$v_{\chi}$ are given by
\begin{eqnarray}
\delta \rho_{\chi} &=& e^{-4 \alpha_0} \chi(t,x^i) \nonumber\\
\delta p_{\chi} &=& \frac{1}{3} \left(1+
\frac{\ddot{\alpha}_0}{\alpha_1^2} \right)e^{-4 \alpha_0}
\chi(t,x^i)
\nonumber\\
(\rho+p)e^{\alpha_0} 
\nabla^2 v_{\chi} &=& e^{-2 \alpha_0} \dot{\chi}(t,x^i)  \ .
\end{eqnarray}
Here, $\chi$ satisfies
\begin{equation}
\ddot{\chi}+ \dot{\alpha}_0 \left(1 -
\frac{\ddot{\alpha}_0}{\alpha_1^2} \right) \dot{\chi}-
\frac{1}{3}\left(1+\frac{\ddot{\alpha}_0}{\alpha_1^2}
\right) e^{- 2 \alpha_0} \nabla^2 \chi =0 \ .
\end{equation}
 At low energies, the equation for $\chi$  becomes
\begin{equation}
 \frac{\partial^2 \chi }{\partial \eta^2} 
      - \frac{1}{3} \nabla^2 \chi=0 \ .
\end{equation}
At large scales and at low energies, $\delta \rho_{\chi}$ is given by
$
\delta \rho_{\chi}=C e^{-4 \alpha_0}
$
where $\chi=C=const.$ 
Thus $\delta \rho_{\chi}$ can be regarded as the perturbations of 
the energy density of the dark radiation. Schematically, we have
$
  \delta \rho = \fbox{standard} + \fbox{$\Phi_0 \neq \Psi_0$}
                   + \fbox{$\delta \rho_\chi$ }
$.
 We can not determine the corrections $\Phi_0 \neq \Psi_0$
 and $\delta \rho_\chi$ from the brane point of view.
 Hence, the bulk analysis is necessary.


\section{Cosmological Perturbations: the view from the bulk}

\subsection{ Bulk Geometry in Poincare coordinate}

The perturbed AdS spacetime in Poincare coordinate system
 is represented by 
\begin{equation}
ds^2= \left(\frac{l}{z} \right)^2
\left(dz^2 - (1+2 \phi) d \tau^2 
+2 b_{,i} dx^i d \tau
+ \left((1 - 2 \hat{\Psi}) \delta_{ij}+ 
2 \hat{E}_{,ij} \right)dx^i dx^j\right) \ . 
\end{equation}
Using this coordinate system, 
the Einstein equation in the bulk becomes 
\begin{equation}
\frac{\partial^2 h}{\partial z^2} -\frac{3}{z} \frac{\partial
h}{\partial z}-\frac{\partial^2 h}{\partial \tau^2}+ \nabla^2 h=0
\end{equation}
where $h=\phi, b, \hat{\Psi}$ and $\hat{E}$.
Here we used the transverse traceless gauge conditions.
The solutions are given by
\begin{eqnarray}
h = \left( \frac{z}{l}\right)^2 \int \frac{d^3 \k}{(2 \pi)^3}
\int d m \:\:
h (m,\k)  Z_2 (m z) e^{-i \omega \tau}e^{i \k \x} \ , \quad
(h= \phi,b,\hat{\Psi},\hat{E})  
\end{eqnarray}
and
\begin{eqnarray}
\phi (m,\k) &=& \frac{2 \k^4}{3 m^2} l^2  E(m,\k) \nonumber\\
b (m,\k) 
&=& -4 i \frac{\sqrt{\k^2+m^2} \:\: \k^2 l^2}{3 m^2} 
E (m,\k) \nonumber\\
\hat{\Psi}(m,\k) &=& 
-\frac{\k^2 l^2}{3}  E (m,\k), \nonumber\\
\hat{E} (m,\k)&=& \frac{2 \k^2 +3 m^2}{3 m^2} l^2 E(m,\k)  \ ,
\end{eqnarray}
where $Z_2$ is the combination of the Hankel function of the
first kind and the second kind of the second rank 
$Z_2(m z)=H^{(1)}_2(m z)+a(m) H^{(2)}_2(m z)$ and  
$\omega^2=m^2+\k^2$. Note that the coefficients 
$E(m,\k)$ and $a(m)$ completely determines the bulk geometry.

\subsection{Bulk geometry in the cosmological coordinate}

The perturbations in the cosmological coordinates  can be
obtained by performing the coordinate transformation;
\begin{equation}
z = z(y,t)=l(f(u)-g(v))=l e^{-\alpha(y,t)} \ , \quad
\tau =\tau(y,t)=l(f(u)+g(v)) \ .
\end{equation}
The  transformation rule is readily obtained as
\begin{equation}
\frac{\partial \tau}{\partial y} 
  = l \dot{\alpha} e^{-\alpha} \ , \ 
\frac{\partial z}{\partial y} =
-l \alpha' e^{-\alpha} \ , \ 
\frac{\partial \tau}{\partial t}
 = l \alpha'  e^{-\alpha} \ , \ 
\frac{\partial z}{\partial t} 
 = -l \dot{\alpha}  e^{-\alpha}   
\end{equation}
After the coordinate transformation, the resulting metric becomes 
\begin{eqnarray}
ds^2 &=&  e^{2 \beta(y,t)} \left((1+2 \hat{N}) dy^2 -(1+2 \hat{\Phi})
d t^2 +2 \hat{A} \: dt \: dy \right)        \\
&& +  e^{2 \alpha(y,t)}
\left(\left((1 - 2 \hat{\Psi}) \delta_{ij}+ 2 \hat{E}_{,ij}\right) 
dx^i dx^j+2 \hat{B}_{,i} dx^i dt + 2 \hat{G}_{,i} dx^i dy \right)  
                                \ ,  \nonumber
\end{eqnarray}
where
\begin{eqnarray}
\hat{\Phi} &=&  (l \alpha')^2 e^{-2 \beta} \phi \nonumber\\
\hat{B} &=&  (l \alpha') e^{-\alpha} b \nonumber\\
\hat{N} &=& - (l \dot{\alpha})^2 e^{-2 \beta} \phi \nonumber\\
\hat{A} &=& -2 (l^2 \dot{\alpha} \alpha') e^{-2 \beta} \phi \nonumber\\
\hat{G} &=& (l \dot{\alpha}) e^{- \alpha} b  \ .
\end{eqnarray}
Note that $\hat{\Psi}, \hat{E}$ does not change.
Thus we have determined the bulk geometry completely.
 What remains is to impose the junction condition at the brane.

\subsection{Cosmological perturbations via projection}

By the gauge transformations, 
$
x^M \to x^M + \xi^M, \quad \xi^M=(\xi^y,\xi^t,\xi^{,i})
$ with 
\begin{equation}
\xi^t = \int^y_0 dy (\hat{A}+ \dot{\xi}^y) + \hat{T}_0 \ , \quad 
\xi = -\int^y_0 dy (\hat{G}+e^{2(\beta-\alpha)} \xi^y) -\hat{E}_0   
\end{equation}
and $\hat{T}_0 = e^{2 \alpha_0} (\hat{B}_0-\dot{\hat{E}}_0)$, 
we can take the Gaussian normal coordinate, $G=A=0$. Note that 
  the brane is located at $y=0$  and we took $B_0=0,E_0=0$ gauge.  
Then we obtain the metric perturbations on the brane as
\begin{equation}
\Phi_0 = \hat{\Phi}_0+ 
\beta_1 \xi^y_0 +\dot{\hat{T}}_0  \ , \  
\Psi_0 = \hat{\Psi}_0- \alpha_1 \xi_0^y -\dot{\alpha}_0 \hat{T}_0 \ , \ 
N_0 = \hat{N}_0 + \xi^{y}_1 + \beta_1 \xi^y_0  
\end{equation}
and the first derivative of the metric perturbations are
\begin{eqnarray}
\Phi_1 &=& \ddot{\xi}^y_0 + \beta_1 \xi^{y}_1+ \beta_2 \xi_0^y 
+ \hat{\Phi}_1 + \dot{\hat{A}}_0+ \dot{\beta}_1 \hat{T}_0
  \ , \nonumber\\
\Psi_1 &=& - \alpha_1 \xi^{y}_1-\dot{\alpha}_0 \dot{\xi}^y_0-\alpha_2 \xi^y_0
+ \hat{\Psi}_1 - \dot{\alpha}_0 \hat{A}_0 -\dot{\alpha}_1 \hat{T}_0
 \ , \nonumber\\
N_1 &=&  \xi^{y}_2+ \beta_1 \xi^y_1 + \beta_2 \xi^y_0 +
\hat{N_1}+ \dot{\beta}_1 \hat{T}_0 
      \ ,\nonumber\\
B_1 &=&  e^{-2 \alpha_0}(-2 \dot{\xi}^y_0 +2 \dot{\alpha}_0 \xi^y_0
- 2 (\beta_1-\alpha_1) \hat{T}_0 
- \hat{A}_0 +e^{2 \alpha_0} \hat{B}_1 - e^{2 \alpha_0} \dot{\hat{G}}_0
)    \ , \nonumber\\
E_1 &=& \hat{E}_1 -e^{-2 \alpha_0} \xi^y_0 -\hat{G}_0 \ .
\nonumber
\end{eqnarray}
And finally junction condition gives 
\begin{eqnarray}
\kappa^2 \delta \rho &=& -6 \left( \dot{\alpha}_0 \dot{\xi}^y_0 
+(\alpha_2 -\alpha_1 \beta_1) \xi^y_0 
-\hat{\Psi}_1 + \dot{\alpha}_0 \hat{A}_0 + \dot{\alpha}_1 \hat{T}_0
-\alpha_1 \hat{N}_0 \right) \ , \nonumber\\
\kappa^2 \delta p &=& 2 \left( \ddot{\xi}_0^y + 2 \dot{\alpha}_0 
\dot{\xi}_0^y 
+(2 \alpha_2 + \beta_2 -\beta_1^2 -2 \alpha_1 \beta_1)\xi_0^y 
\right. \nonumber\\
&& \!\!\!\!\!\! \left. + \hat{\Phi}_1 -2 \hat{\Psi}_1 + \dot{\hat{A}}_0
+2 \dot{\alpha}_0 \hat{A}_0 
+(\dot{\beta}_1+2 \dot{\alpha}_1) \hat{T}_0 
- (\beta_1+2 \alpha_1)\hat{N}_0 \right) \ , \nonumber\\
\kappa^2(\rho+p) e^{\alpha_0} v &=& 2 \dot{\xi}_0^y -2 \dot{\alpha}_0 \xi^y_0 
- e^{2 \alpha_0} \hat{B}_1 
+ 2 (\beta_1-\alpha_1) \hat{T}_0 +e^{2 \alpha_0} \dot{\hat{G}}_0 + \hat{A}_0 
                       \ ,\nonumber\\
0&=& -2 e^{-2 \alpha_0} \xi^y_0 +2 \hat{E}_1 -2 \hat{G}_0 \ .
\end{eqnarray}
From the last equation, $\xi^y_0$ is determined by
$E(m,\k)$ and $a(m)$. 
 Thus, the quantities on the brane are determined by the bulk
 geometry.

\subsection{Results}

Substituting the explicit solutions (5.4) and (5.8)
 into eq. (5.11),  we obtain
\begin{eqnarray*}
-\frac{\kappa^2 \alpha_1}{2} \delta \rho (\k)
&=& -3 (\dot{\alpha}_0 \dot{\Psi}_0 + \dot{\alpha}_0^2 \Phi_0)
-e^{-2 \alpha_0} \k^2 \Psi_0       \\
&&  +  
\frac{1}{3}e^{-4 \alpha_0} \int dm E(m,\k) \k^4 l^2
Z_0(m l e^{-\alpha_0}) e^{-i \omega T(t)}     \ ,
\end{eqnarray*}
\begin{eqnarray*}
-\frac{\kappa^2 \alpha_1}{2} \delta p(\k)
&=& \ddot{\Psi}_0+\left(3 \dot{\alpha}_0-\frac{\dot{\alpha_0} 
\ddot{\alpha}_0}{\alpha_1^2}\right) \dot{\Psi}_0+
\dot{\alpha}_0 \dot{\Phi}_0 +\left( 2 \ddot{\alpha}_0
-\frac{\dot{\alpha}_0^2 \ddot{\alpha}_0}{\alpha_1^2} +3 \dot{\alpha}_0^2
\right) \Phi_0     \\
&-& \frac{1}{3}e^{-2 \alpha_0} \k^2 \Phi_0+\frac{1}{3}
\left(1-\frac{\ddot{\alpha}_0}{\alpha_1^2} \right)
e^{-2 \alpha_0}\k^2 \Psi_0    \\
&+& \frac{1}{9}
\left(1+\frac{\ddot{\alpha}_0}{\alpha_1^2} \right)
e^{-4 \alpha_0} \int dm E(m,\k) \k^4 l^2
Z_0(m l e^{-\alpha_0}) e^{-i \omega T(t)}  \ ,
\end{eqnarray*}
\begin{eqnarray}
&&-\frac{\kappa^2 \alpha_1}{2} (\rho+p) e^{\alpha_0}v(\k) 
 =\dot{\Psi}_0 +\dot{\alpha}_0 \Phi_0  \nonumber\\
&& \qquad + \frac{1}{3} e^{-3\alpha_0} \int dm E(m,\k)  
\left( 
\alpha_1 i \omega \k^2 l^3 Z_0 
(m l e^{-\alpha_0}) -\dot{\alpha}_0 
m \k^2 l^3  Z_1(m l e^{-\alpha_0})
\right) e^{-i \omega T(t)} \ ,
\end{eqnarray}
where we denoted $\tau(0,t)=T(t)$. 
Using eq. (5.10), we can also obtain the metric perturbations 
 as 
\begin{eqnarray}
\Psi_0(\k) &=& \int dm E(m,\k)\left(
m l e^{-\alpha_0} Z_1 (m l e^{-\alpha_0})
+ \frac{1}{3} (\k l e^{-\alpha_0})^2 Z_0(m l 
e^{-\alpha_0})\right) e^{-i \omega T(t)} \ ,\nonumber\\
\!\! \Phi_0(\k)&=& 
\int dm E(m,\k) 
\left(
m l e^{-\alpha_0} Z_1 (m l e^{-\alpha_0})
-\frac{1}{3}(\k^2+3 m^2)l^2 e^{-2 \alpha_0}
Z_0(m l e^{-\alpha_0}) \right) e^{-i \omega T(t)}
\nonumber\\
&& \!\!\!\!\!\!\!\!\!\!\!\!\!\!\!\!\!\!\!\!\!
+(\dot{\alpha}_0 l)^2 \int dm E(m,\k)
\left(mle^{-\alpha_0} Z_1(m l e^{-\alpha_0}) 
- (\k^2+2 m^2)l^2  e^{-2 \alpha_0}
Z_0(m l e^{-\alpha_0}) \right)e^{-i \omega T(t)} \nonumber\\
&& -2 \alpha_1 \dot{\alpha}_0 l^2 \int dm E(m,\k) 
(i \omega m l^2 e^{-2 \alpha_0}) Z_1(m l e^{-\alpha_0}) 
e^{-i \omega T(t)}  \ .
\end{eqnarray} 
 Now we are in a position to discuss the consequence of the
  bulk analysis.

\section{ Conclusion } 
 
Let us consider the long wave-length perturbations: 
$ \k l e^{-\alpha_0} \to 0  \ ,\ \ \ 
\k \dot{\alpha}_0^{-1} e^{-\alpha_0} \ll 1 $. 
In this case, the bulk solution combined with the information
 from the brane point of view gives the useful results.

Now we shall start with the discussion of the 
 $\delta \rho_\chi$.
 At low energies $\dot{\alpha}_0 l \ll 1$, in case that  
$E(m,\k)$  have a peak at $2 \k^2+3 m^2=0$, i.e. 
$E^{(\chi)}(\k)=E(m_k=\sqrt{2/3} \k i,\k)$, 
the metric perturbations become
\begin{equation}
\Psi_0=\Phi_0= \frac{1}{2}E^{(\chi)}(\k) (m_k l e^{- \alpha_0})^2
Z_2(m_k l e^{-\alpha_0}) e^{\frac{1}{\sqrt{3}}i \k \eta} \ ,
\end{equation}
where we used $T=-\eta$ at low energies and $\eta$ is
the conformal time. 
If we take 
\begin{equation}
a(m)= a^{(\chi)}(m)=-\frac{H^{(1)}_2(m_k l e^{-\alpha_0})}
{H^{(2)}_2 (m_k l e^{-\alpha_0})} \ ,
\end{equation}
we get $\Phi_0=\Psi_0=0$. 
Then, the density perturbations $\delta \rho$ becomes
\begin{equation}
\frac{\kappa^2}{2 l}
\delta \rho
=  \frac{1}{3} e^{-4 \alpha_0} \k^4 l^2 
E^{(\chi)}(\k) Z^{(\chi)}_0 (m_k l e^{-\alpha_0}) 
e^{\frac{1}{\sqrt{3}} i \k \eta} \ ,
\end{equation}
where $Z^{(\chi)}_0=H^{(1)}_0+a^{(\chi)}(m) H^{(2)}_0$.
Because these perturbations do not contribute to the 
metric perturbations and $\delta \rho$
 satisfies
\begin{equation}
\delta \rho = e^{-4 \alpha_0} \chi ,
\quad \chi'' +\frac{1}{3} \k^2 \chi=0 ,
\end{equation}
they should be identified with $\delta \rho_{\chi}$.
 They diverge at the horizon of the AdS spacetime
as $\exp(\sqrt{2/3} \k z)$ as $z \to \infty$. 
Thus if we impose the regularity on the perturbations 
 in the bulk, the corrections 
from $\chi$ should not exist on the brane, 
$
\delta \rho_{\chi}=\delta p_{\chi}=v_{\chi}=0
$

Next, we will determine $\Phi_0 - \Psi_0 $. 
The metric perturbations are given by
\begin{eqnarray}
\Phi_0 &=& (1+(\dot{\alpha}_0 l)^2)\Psi_0 
 -(1+2 (\dot{\alpha}_0 l)^2)
\int dm E^{(1)}(m) 
(m l e^{-\alpha_0})^2 Z_0(ml e^{-\alpha_0})e^{-i m T} 
 \nonumber \\
&& -  2 i \alpha_1 \dot{\alpha}_0 l^2 
\int dm E^{(1)}(m)(m l e^{-\alpha_0})^2  
Z_1(m l e^{-\alpha_0})e^{- i m T}  \ .
\end{eqnarray}
 We will assume that only the modes with $m le^{-\alpha_0}
 \ll 1$ can contribute to the perturbations in the bulk.
Then using the asymptotic form of the Hankel function, 
we obtain
\begin{equation}
\Phi_0 = (1+ (\dot{\alpha}_0 l)^2) \Psi_0 \ .
\end{equation}
At low energies $\dot{\alpha}_0 l \ll 1$, we have
$
\Psi_0=\Phi_0
$.
Then the metric perturbations are obtained as
\begin{equation}
\Phi_0 = \Psi_0=\frac{3(1+w)}{5+3 w} \zeta_{\ast} \ , \quad
\frac{\delta \rho}{\rho} = -2 \Phi_0  \ .
\end{equation} 
 At high energies, we have
$
\Phi_0 = (\dot{\alpha}_0 l)^2 \Psi_0
$.
then $\Phi_0 \gg \Psi_0$.
we get
\begin{equation}
\Phi_0 = 3(1+w) \zeta_{\ast}   \ ,  \quad
\frac{\delta \rho}{\rho} = - \Phi_0  \ .
\end{equation}

 We have attempted to impose a natural boundary condition, i.e.
  regularity at the horizon. And we assumed that the massive modes
  can be neglected. Under these conditions, we have successfully
  obtained the complete information to determine the evolution of
  the cosmological perturbations on the brane. In doing that, 
  both the brane and bulk analysis have taken important roles.

\section*{Acknowledgments}
We would like to thank the participants of the YITP workshop
 YITP-W-01-15 on ``Braneworld-Dynamics of spacetime boundary"
  for fruitful discussions. This work was supported in part by
  Monbukagakusho Grant-in-Aid No.14540258.

\end{document}